\documentclass[12pt,a4paper]{article}

\usepackage{amsmath}
\usepackage{bbold}
\usepackage{amsfonts}
\usepackage{amssymb}
\usepackage{graphicx, rotating}
\usepackage{epstopdf}
\usepackage{epsfig}
\usepackage{latexsym}
\usepackage{graphicx}
\usepackage{color}
\usepackage{cite}
\usepackage{slashed}
\usepackage{hyperref}

\usepackage{epsfig}\parskip 5pt plus 1pt
\usepackage{bbm}
\usepackage{amsmath}
\usepackage{amssymb}
\usepackage{amsfonts}
\usepackage{mathrsfs}
\usepackage{graphicx}
\usepackage{axodraw}
\usepackage{color}
\usepackage{bbold}

\begin{document}

\begin{titlepage}
\vspace{.3in}
\flushright ZU-TH 40/14
\begin{center}
\vspace{1cm}

{\Large \bf
Minimal muon anomalous magnetic moment\\
}

\vspace{1.2cm}

{\large Carla Biggio$^{a}$ and Marzia Bordone$^{b}$\\}
\vspace{.8cm}

{\it $^a$Dipartimento di Fisica, Universit\`a di Genova \& INFN,
  Sezione di Genova, \\
via Dodecaneso 33, 16159 Genova, Italy }\\
{\it $^b$Physik-Institut, Universit\"{a}t Z\"{u}rich,\\
 Wintherturerstrasse 190, CH-8057 Z\"{u}rich, Switzerland }\\


\end{center}
\vspace{.8cm}

\begin{abstract}
\medskip
\noindent
We classify all possible one-particle (scalar and fermion) extensions
of the Standard Model that can contribute to the anomalous magnetic
moment of leptons. We review the cases already discussed in the literature and complete the picture
by performing the calculation for a
fermionic doublet with hypercharge $-3/2$. We conclude that, out of
the listed possibilities,
only two scalar leptoquarks and the pseudoscalar of a peculiar
two-Higgs-doublet model could be the responsibles for the muon anomalous
magnetic moment discrepancy. Were this the case, this particles
could be seen in the next LHC run. To this aim, especially to test the
leptoquark hypothesis, we suggest to look for final states with tops
and muons.
\end{abstract}


\end{titlepage}

\section{Introduction} 

The anomalous magnetic moments of electron and muon are among
the best measured quantities ever. With its 12 digits, the measurement of
the anomalous magnetic moment of the electron is used to fix the value
of the fine structure constant $\alpha_{em}$, while the muon anomalous
magnetic moment, $a_\mu = \frac{g-2}{2},$ can be used
to test the Standard Model (SM) or, in other words,
to search for new physics. 

In 2006 the experiment E821 carried out at the Brookhaven laboratories measured~\cite{Bennett:2006fi}
\begin{equation}
a_\mu^{exp} = 116592080 (63) \cdot 10^{-11}\ ,
\end{equation}
while the value predicted within the SM is given by~\cite{Jegerlehner:2009ry}
\begin{equation}
a_\mu^{SM} =  116591790 (65) \cdot 10^{-11} \ .
\end{equation}
The difference between the predicted and measured value, $\Delta a_\mu
= 290(90)  \cdot 10^{-11}$, constitutes a discrepancy with
3.1$\sigma$ significance.

The SM result is not univocal, since the hadronic contribution
depends on some experimental inputs. The
above quoted result is obtained by taking data from
$e^+ e^-$ annihilation. If, on the other hand, one uses $\tau$-decay data, a slightly
higher value is obtained~\cite{Davier:2003pw}, leading to a smaller
$\Delta a_\mu$. Therefore one could think that the
reason of such a discrepancy is due to a poor knowledge of the
theoretical calculation. However, if one fixes by hand the hadronic contribution in order to obtain the experimental result, other SM predictions turn out
to be affected, in such a
way that this possibility is substancially
excluded~\cite{passeraetal}. Therefore, the reason of the
discrepancy seems not to lie in the SM calculation.

On the other hand, the measurement of the $(g-2)_\mu$ has been
performed only by one experiment, so that the possibility exists that
some systematics out of control affect the result. In order to have a
confirmation or a disproof of it, the experiment E989 is under construction
at Fermilab~\cite{fermilab-exp} and another one has been proposed in
Japan, at J-PARC~\cite{jparc}.

Therefore, if one takes the
experimental result as firm as well as the SM calculation, 
the only possibility to explain the discrepancy is to
invoke the presence of new physics (NP). In the following we will make this
assumption and analyse which kind of NP could be there.

The contribution of new physics to the muon anomalous magnetic moment
has been considered in many extensions of the SM. For example, in
supersymmetric (SUSY) models it is quite easy to get the needed
$\Delta a_\mu$, when two sparticles circulate
in the loop. However, this is not the
only possibility and, in particular, it is possible to give a positive
contribution to $a_\mu$ with the simple addition of one new particle.
In this paper we focus precisely on this hypothesis and
classify all possible one-particle SM extensions that can give rise,
at one loop, to a contribution to the anomalous magnetic moment of the
muon. We consider only scalars and fermions. Indeed the addition of
new massive vector bosons would imply the extension of the SM
gauge group, a possibility that we do not want to consider here. Most of the cases that
we will find in our general classification have been already studied
in the literature. Here we aim to complete the picture by calculating
the contribution to $a_\mu$ given by the addition to the SM of a fermionic
colourless SU(2) doublet with hypercharge $-3/2$, a result which is not present
in the literature. Afterwards, we
briefly review the literature regarding the other cases of the list,
comment and update some results, and draw our conclusions.

\section{The classification}

In the SM the anomalous magnetic moment of leptons is generated through the
dimension-six effective operator
\begin{equation}
\frac{1}{\Lambda^2} \bar{L} \sigma^{\mu\nu} e_R \phi F_{\mu\nu} +
h.c. \ ,
\end{equation}
where $L$ is the lepton doublet, $e_R$ the lepton singlet, $\phi$ the
Higgs doublet and $F_{\mu\nu}$ the electromagnetic field strength.
This is generated at one loop, with a fermion and a gauge or Higgs
boson circulating in it. In SM extensions, an analogous loop, with a
new fermion or boson inside it, can give a contribution to the same
operator. Here {\it we want to  classify all one-particle extensions of the SM that can
contribute to the anomalous magnetic moment of leptons}. As discussed
before, we only look for scalars and fermions circulating in the
loop.
The only thing we have to require is that at any vertex Lorentz
invariance, gauge invariance and renormalizability are
respected.~\footnote{A non-gauge-invariant approach have been adopted
  in Ref.~\cite{Queiroz:2014zfa}.} This gives us a finite list of new
particles, that are collected in tables~\ref{tab:fermioni}
and~\ref{tab:scalari}~\footnote{In our notation the Higgs doublet $\phi$
has hypercharge $+1/2$.}. Some comments
are in order.

\begin{table}[!t]
\begin{equation*}
\begin{array}{ccccc}
\hline
 & SU\left(3\right) & SU\left(2\right) & Y & Q\\
\hline
N_{\scriptscriptstyle R} & 1 & 1 & 0 & 0 \\[0,6mm]
\Sigma_{\scriptscriptstyle R} & 1 & 3 & 0 & 1,0,-1 \\[0,6mm]
E_{\scriptscriptstyle 4} & 1 & 1 & -1 & -1\\[0,6mm]
L_{\scriptscriptstyle 4} & 1 & 2 & -\frac{1}{2} & 0,-1\\[0,6mm]
T & 1 & 3 & -1 & 0,-1,-2\\[0,6mm]
D & 1 & 2 & -\frac{3}{2} & -2,-1\\[0,6mm]
\hline
\end{array}
\end{equation*}
\caption{New fermion fields with their quantum numbers.}
\label{tab:fermioni}
\end{table}
\begin{table}[!t]
\begin{equation*}
\begin{array}{ccccc}
\hline
 & SU\left(3\right) & SU\left(2\right) & Y &
Q\\
\hline
S_{1} & 1 & 1 & 1 & 1 \\[0,6mm]
S_{2} & 1 & 1 & 2 & 2\\[0,6mm]
H_{2} & 1 & 2 & \frac{1}{2} & 1,0\\[0,6mm]
\Delta & 1 & 3 & 1 & 2,1,0 \\[0,6mm]
T_{\scriptscriptstyle c}^{\scriptscriptstyle 1/3} & \bar{3} & 3 & \frac{1}{3} & \frac{4}{3},\frac{2}{3},\frac{1}{3}\\[0,6mm]
S_{\scriptscriptstyle c}^{\scriptscriptstyle 1/3} & \bar{3} & 1 & \frac{1}{3} & \frac{1}{3}\\[0,6mm]
S_{\scriptscriptstyle c}^{\scriptscriptstyle 4/3} & \bar{3} & 1 & \frac{4}{3} & \frac{4}{3}\\[0,6mm]
D_{\scriptscriptstyle c}^{\scriptscriptstyle 7/6} & 3 & 2 & \frac{7}{6} & \frac{5}{3},\frac{2}{3}\\[0,6mm]
D_{\scriptscriptstyle c}^{\scriptscriptstyle 1/6} & 3 & 2 &
\frac{1}{6} & \frac{2}{3},\frac{1}{3}\\[0,6mm]
\hline
\end{array}
\end{equation*}
\caption{New scalar fields with their quantum numbers.}
\label{tab:scalari}
\end{table}

\subsection{Fermions}

As for new fermions, we have obtained them by looking for fermions
circulating in the loop together with a Higgs boson. If on the
external legs there are the lepton doublets, one gets new fermions
which are colour singlets, SU(2) singlets or triplets, with 0 or 1
hypercharge. On the other hand, if on the external legs there are
lepton singlets, one gets new fermions which are colourless, doublets
under SU(2), with hypercharge either $-1/2$ or $-3/2$. 

In principle one could think of looking for new fermions circulating
in the loop together with gauge bosons. For example, if the circulating gauge
bosons are the SU(2) ones, a
new field transforming in the 4 of SU(2) would be obtained. However such kind of interaction is
non-renormalizable, since it is given by the operator
\begin{equation}
\label{NRint}
\bar{\psi} \sigma_{\mu\nu} \tau^A \chi W_A^{\mu\nu} \, ,
\end{equation}
where $\chi$ is the new field and $\psi$ a SM fermion.
Therefore, we will not consider new fields obtained in this way, since
they correspond to non-minimal extensions of the SM.

All the fermions we consider, except for the neutral ones, have to be
Dirac fermions. This guarantees on the one side the presence of a
Dirac mass, independent from the electroweak (EW) symmetry breaking, that can therefore assume any
value and, on the other hand, that our extension of the SM is
anomaly-free. Moreover, in the list of table~\ref{tab:fermioni}, we
recognise some well-known field: $N_R$ is the right-handed
neutrino, which can give mass to a light neutrino via a type-I seesaw
mechanism~\cite{typeI}; $\Sigma_R$ is a SU(2) triplet that can realize
the type-III seesaw~\cite{typeIII}; $L_4$ and $E_4$ are copies of the
lepton doublets and singlets, but vector-like; $T$ is a triplet, whose
phenomenology have been studied in Ref.~\cite{Delgado:2011iz}; $D$ is
a doublet recently discussed in Ref.~\cite{Ma:2014zda}. Actually
these are the only fermion fields which mix with leptons at tree level
respecting the SM gauge invariance; their effect on EW observable
have been studied in Ref.~\cite{delAguila:2008pw}.

\subsection{Scalars}

In the case of new scalars, more possibilities arise, depending on the fields
on the external legs and inside the loop. If only lepton doublets are
involved, one obtain colourless scalars with hypercharge 1, singlets
or triplets of SU(2). If only lepton singlets are there, the scalar
must be colourless, SU(2) singlet, with hypercharge 2. If both lepton
singlets and doublets are considered, the only possibility is a
replica of the Higgs doublet, with quantum numbers $(1,2,1/2)$. 

On
the other hand, if quarks circulate in the loop, the new scalars must be
leptoquarks, i.e. coupling to both quarks and leptons. When there are
lepton doublets on the external legs and a quark doublet in the loop,
the new scalar transforms in the $\bar{3}$ of SU(3), can be either a
singlet or a triplet of SU(2) and has hypercharge 1/3. This singlet
is obtained also with only singlets around, together with a similar
one with hypercharge 4/3. On the contrary, with doublets on the
external legs and quark singlets inside the loop one obtain new
scalars transforming as SU(3) triplets, SU(2) doublets and with
hypercharge 1/6 or 7/6. This last is obtained also with leptons singlets in the external
legs and a quark doublet inside. All these leptoquarks have been
already classified and considered to solve the $(g-2)_\mu$ problem in Ref.~\cite{Chakraverty:2001yg}.

\section{The contribution of the colourless SU(2) doublet with
  hypercharge $-3/2$}

In this section we calculate the contribution to the muon anomalous
magnetic moment given by extending the SM with a colourless
SU(2) fermion doublet, with hypercharge $-3/2$, named $D$. This field
is composed by two charged particles, named $\chi$ and $\Psi$, with
charges equal to $-1$ and $-2$, respectively:
\begin{equation}
D=\left(
\begin{array}{c}
\chi \\ \Psi 
\end{array}
\right) \, .
\label{eq:D}
\end{equation}
The SM Lagrangian has to be enlarged to include kinetic, mass and
interaction terms for $D$:
\begin{equation}
\label{Lagrangian}
\mathcal{L}=
\mathcal{L}_{SM}+\bar{D} (\slashed{\mathcal D}-M_D) D-(\lambda_D\bar{D}_L\phi^c
l_R+h.c.) \, ,
\end{equation}
where $\lambda_D=(\lambda_{De}, \lambda_{D\mu}, \lambda_{D\tau})$ are the
Yukawa couplings between $D$ and the SM leptons, $\phi^c = i\sigma_2
\phi^*$, $M_D$ is the Dirac mass of the new field
  and $\mathcal{L}_{SM}$ is the SM Lagrangian.

Due to the Yukawa interaction, the singly charged component of $D$,
$\chi$, mixes with the SM charged leptons. In the interaction basis their mass matrix is indeed given by 
\begin{equation}
\label{Mass}
-\begin{pmatrix}
\bar{\ell}_L & \bar{\chi}_L
\end{pmatrix} 
\begin{pmatrix}
\frac{yv}{\sqrt{2}} & 0 \\[1.5mm]
\frac{\lambda_D v}{\sqrt{2}} & M_D
\end{pmatrix} 
\begin{pmatrix}
\ell_R \\[1mm]
\chi_R
\end{pmatrix} 
+h.c. \ ,
\end{equation}
where $\ell = (e\ \mu\ \tau)^T$, $y$ is the diagonal matrix of the SM leptonic Yukawa
couplings and $v=246$~GeV. After the diagonalisation of the mass matrix and the
consequent fields redefinition, the interaction Lagrangian gets
modified and the novel contribution to the anomalous magnetic moment of
the muon can be calculated. We refer to the Appendix for the details
on the diagonalization procedure and the couplings in the mass basis. 

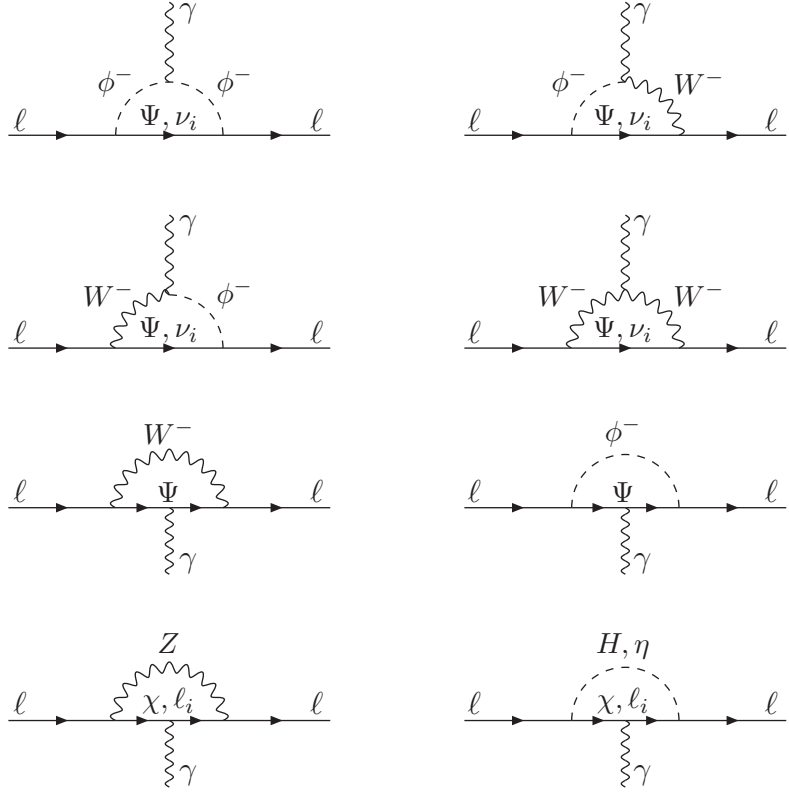
\begin{figure}[t]
\centering
\vspace{50mm}
\begin{picture}(330,150)(0,-175)
\ArrowLine(10,80)(50,80)
\ArrowLine(50,80)(90,80)
\ArrowLine(90,80)(130,80)
\DashCArc(70,80)(20,0,180){3}
\Photon(70,100)(70,130){1.5}{5}
\Text(12,82)[bl]{$\ell$}
\Text(128, 82)[br]{$\ell$}
\Text(74,130)[tl]{$\gamma$}
\Text(57,95)[br]{\small{$\phi^-$}}
\Text(88,95)[bl]{\small{$\phi^-$}}
\Text(70,82)[b]{\small{$\Psi , \nu_i$}}
\ArrowLine(180,80)(220,80)
\ArrowLine(220,80)(260,80)
\ArrowLine(260,80)(300,80)
\DashCArc(240,80)(20,90,180){3}
\PhotonArc(240,80)(20,0,90){2}{5.5}
\Photon(240,100)(240,130){1.5}{5}
\Text(182,82)[bl]{$\ell$}
\Text(298, 82)[br]{$\ell$}
\Text(244,130)[tl]{$\gamma$}
\Text(227,95)[br]{\small{$\phi^-$}}
\Text(259,96)[bl]{\small{$W^-$}}
\Text(240,82)[b]{\small{$\Psi , \nu_i$}}
\ArrowLine(10,0)(50,0)
\ArrowLine(50,0)(90,0)
\ArrowLine(90,0)(130,0)
\DashCArc(70,0)(20,0,90){3}
\PhotonArc(70,0)(20,90,180){2}{5.5}
\Photon(70,20)(70,50){1.5}{5}
\Text(12,2)[bl]{$\ell$}
\Text(128, 2)[br]{$\ell$}
\Text(74,50)[tl]{$\gamma$}
\Text(57,16)[br]{\small{$W^-$}}
\Text(88,15)[bl]{\small{$\phi^-$}}
\Text(70,2)[b]{\small{$\Psi , \nu_i$}}
\ArrowLine(180,0)(220,0)
\ArrowLine(220,0)(260,0)
\ArrowLine(260,0)(300,0)
\PhotonArc(240,0)(20,0,180){2}{10.5}
\Photon(240,22)(240,50){1.5}{5}
\Text(182,2)[bl]{$\ell$}
\Text(298,2)[br]{$\ell$}
\Text(244,50)[tl]{$\gamma$}
\Text(227,16)[br]{\small{$W^-$}}
\Text(259,16)[bl]{\small{$W^-$}}
\Text(240,2)[b]{\small{$\Psi , \nu_i$}}
\ArrowLine(10,-60)(50,-60)
\ArrowLine(50,-60)(70,-60)
\ArrowLine(70,-60)(90,-60)
\ArrowLine(90,-60)(130,-60)
\PhotonArc(70,-60)(20,0,180){2}{10.5}
\Photon(70,-60)(70,-85){1.5}{5}
\Text(12,-58)[bl]{$\ell$}
\Text(128, -58)[br]{$\ell$}
\Text(74,-85)[bl]{$\gamma$}
\Text(70,-36)[b]{\small{$W^-$}}
\Text(70,-58)[b]{\small{$\Psi$}}
\ArrowLine(180,-60)(220,-60)
\ArrowLine(220,-60)(240,-60)
\ArrowLine(240,-60)(260,-60)
\ArrowLine(260,-60)(300,-60)
\DashCArc(240,-60)(20,0,180){3}
\Photon(240,-60)(240,-85){1.5}{5}
\Text(182,-58)[bl]{$\ell$}
\Text(298, -58)[br]{$\ell$}
\Text(244,-85)[bl]{$\gamma$}
\Text(240,-37)[b]{\small{$\phi^-$}}
\Text(240,-58)[b]{\small{$\Psi$}}
\ArrowLine(10,-140)(50,-140)
\ArrowLine(50,-140)(70,-140)
\ArrowLine(70,-140)(90,-140)
\ArrowLine(90,-140)(130,-140)
\PhotonArc(70,-140)(20,0,180){2}{10.5}
\Photon(70,-140)(70,-165){1.5}{5}
\Text(12,-138)[bl]{$\ell$}
\Text(128, -138)[br]{$\ell$}
\Text(74,-165)[bl]{$\gamma$}
\Text(70,-115)[b]{\small{$Z$}}
\Text(70,-138)[b]{\small{$\chi , \ell_i$}}
\ArrowLine(180,-140)(220,-140)
\ArrowLine(220,-140)(240,-140)
\ArrowLine(240,-140)(260,-140)
\ArrowLine(260,-140)(300,-140)
\DashCArc(240,-140)(20,0,180){3}
\Photon(240,-140)(240,-165){1.5}{5}
\Text(182,-138)[bl]{$\ell$}
\Text(298, -138)[br]{$\ell$}
\Text(244,-165)[bl]{$\gamma$}
\Text(240,-117)[b]{\small{$H, \eta$}}
\Text(240,-138)[b]{\small{$\chi , \ell_i$}}
\end{picture}
\caption{Diagrams contributing to the anomalous magnetic moment of the
  lepton $\ell$ (in the Feynman gauge). $\phi^\pm$ and $\eta$ are the three Goldstone bosons
  associated with the $W^\pm$ and $Z$ bosons, while $H$ stands for the
  physical Higgs boson.}
\label{fig:diagrams}
\end{figure}

In order to calculate the contribution of the doublet $D$ to the
$(g-2)_\mu$, the diagrams in Fig.~\ref{fig:diagrams} have to be
calculated. At $\mathcal{O}(\frac{v^2\lambda_D
  \lambda_D^\dagger}{M_D^2})$, we obtain:
\begin{equation}
\label{amuD1}
\begin{split}
a_{\mu}^{\scriptscriptstyle \text{SM}+D}=\frac{m_\mu^{2}
  G_F}{24\sqrt{2}\pi^{2}}\bigg\{&\left(3-4\cos^{2}\theta_ W\right)^{2}+5+ 
\frac{v^{2}\left|\lambda_{D\mu}\right|^{2}}{M_D^{2}}\bigg[-\frac{11}{4}-4\cos^{2}\theta_W +\\
&+F_{\text{NC}}\left(\frac{M_D^{2}}{M_Z^{2}}\right)+F_{\text{h}}\left(\frac{M_D^{2}}{M_H^{2}}\right)
+F_{\text{CC}}\left(\frac{M_D^{2}}{M_W^{2}}\right)\bigg]\bigg\} \ ,
\end{split}
\end{equation}
where
\begin{subequations}
\begin{align}
F_{\text{NC}}\left(k\right)=& \ \frac{16+14k-63k^2+38k^3-5k^5-6\left(2-13k+8k^2\right)\log k}{4\left(k-1\right)^{4}}, \\
F_{\text{h}}\left(k\right)=& \ \frac{7k^4-36k^3+45k^2-16k+6k\left(3k-2\right)\log k}{4\left(k-1\right)^{4}}, \\
F_{\text{CC}}\left(k\right)=& \ \frac{46-79k+42k^{2}-13k^{3}+4k^{4}+6\left(-4+24k-20k^2+3k^{3}\right)\log k}{2\left(k-1\right)^{4}} \ .
\end{align}
\end{subequations}
We recognize, in the $\lambda_D$-independent term, the well-known SM
electroweak contribution~\cite{g-2-EW}. Other terms are the contribution
of the new doublet which depends on $D$ mass and couplings. The terms
in the square brackets are a function of $M_D$, which is negative
in the mass interval allowed, i.e.  for $M_D > 100.8$~GeV
(LEP bounds~\cite{pdg}). Therefore,
{\it the addition of the
doublet $D$ to the SM field content cannot explain the anomaly in
anomalous magnetic moment of the muon}.

Even if this model is not good to solve the discrepancy of the
$(g-2)_\mu$, we can estimate the maximum value of this negative
contribution. The combination
$\frac{v^{2}\left|\lambda_{D\mu}\right|^{2}}{M_D^{2}}$ is indeed
constrained by EW fits~\cite{delAguila:2008pw} to be
\begin{equation}
\frac{v \left|\lambda_{D\mu}\right|}{\sqrt{2} M_D}\leq 0.028 \ ,
\end{equation}
while the maximum value of the modulus of the square bracket is obtained for the
minimum allowed value for the $D$ mass, 100.8~GeV, and it
corresponds to -91.8. By maximizing independently these two quantities
we have an estimation of the maximum correction that the presence of
such a doublet can induce in the anomalous magnetic moment of the
muon:
\begin{equation}
\label{amuDlimit}
\left|a_{\mu}^D \right|\leq 5.5 \times 10^{-11}
\ .
\end{equation}
This value is smaller than the theoretical and the experimental errors
(and thus it will not be relevant anyway in the near future).

\section{Other cases}

\subsection{Fermions}

\noindent $\mathbf{N_R \sim (1,1,0) \quad  \& \quad   \Sigma_R \sim (1,3,0) }$

The contribution to $a_\mu$ of these new particles has been calculated in
Refs.~\cite{Biggio:2008in,Freitas:2014pua}. Despite some minor
differences in their results, the outcome is the same: their
contribution is always negative and, as such, can not generate $\Delta a_\mu$.\\

\noindent $\mathbf{E_4 \sim (1,1,-1) \quad  \& \quad   L_4 \sim (1,2,-1/2) }$

The calculation in these cases has been presented in
Ref.~\cite{Freitas:2014pua}~\footnote{With respect to their results,
  we get a discrepancy in the constant factor in both cases, while
  loop functions and $\theta$-dependent terms are the same:
\begin{eqnarray}
\nonumber
a_\mu^{E_4} &=& \frac{G_F
  m_\mu^2}{16\sqrt{2}\pi^2}\frac{v^2|\lambda_{E\mu}|^2}{M_E^2}
\left [-\frac{8}{3}\cos^2\theta_w
  +\frac{5}{6}+F_{FFV}(\frac{M_L^2}{M_Z^2})+H_{FFS}(\frac{M_L^2}{M_H^2})
\right]\\
\nonumber
a_\mu^{L_4} &=& \frac{G_F
  m_\mu^2}{16\sqrt{2}\pi^2}\frac{v^2|\lambda_{L\mu}|^2}{M_L^2}
\left [ \frac{8}{3}\cos^2\theta_w
  +\frac{5}{6}+F_{FFV}(\frac{M_L^2}{M_Z^2})+H_{FFS}(\frac{M_L^2}{M_H^2})+
2F_{VVF}(\frac{M_L^2}{M_W^2})+2G_{VVF}(\frac{M_L^2}{M_W^2})
\right]\ .
\end{eqnarray}
We have adopted the notation of Ref.~\cite{Freitas:2014pua} for the loop functions.
We have also cross-checked our results with Ref.~\cite{Chiu:2014oma} and
concluded that the discrepancy could be due to a difference in
the result of the calculation of the $Z$-diagram. According to our
results, the contribution of $E_4$ is always negative, while the
contribution of $L_4$ is positive for masses above 119~GeV. Anyway,
this difference does not modify the substance of the conclusion.}. 
In the $L_4$ case, the contribution to $a_\mu$ is positive, at least for masses
above 119.3~GeV. This is the only case where the addition of a single
fermion gives a positive contribution. However, as before,
there are bounds on the combination
$\frac{v^{2}\left|\lambda_{L\mu}\right|^{2}}{M_L^{2}}$ coming from EW fits~\cite{delAguila:2008pw}:
\begin{equation}
\frac{v \left|\lambda_{L\mu}\right|}{\sqrt{2} M_D}\leq 0.048 \ .
\end{equation}
This implies the following upper bound on the contribution to $a_\mu$
\begin{equation}
a_\mu^{L_4} < 6.8 \cdot 10^{-12}\ ,
\end{equation}
which turns out to be much smaller than what needed to solve the
discrepancy between the theoretical and experimental value of
$a_\mu$.\footnote{On the contrary, notice that, as it has been shown in Refs.~\cite{Kannike:2011ng,Dermisek:2013gta},
  the simultaneous addition of $E_4$ and $L_4$ can solve the muon
  $(g-2)$ puzzle.}\\

\noindent $\mathbf{T \sim (1,3,-1)}$

As for the contribution of the SU(2) triplet with hypercharge $-1$,
it has been calculated in Ref.~\cite{Freitas:2014pua}: also in this
case the
contribution is negative, independently of the mass of the new
particle, and cannot therefore explain the discrepancy.

\subsection{Scalars}

\noindent $\mathbf{S_1 \sim (1,1,1)}$

The contribution to $a_\mu$ in this case has been calculated in
Ref.~\cite{CoarasaPerez:1995wa} and can also be derived from the results obtained more
recently in Ref.~\cite{Chiu:2014oma}. It turns out to be always negative and, as
such, cannot solve the $a_\mu$ problem.\\

\noindent $\mathbf{S_2 \sim (1,1,2)}$

In this case the contribution is given by two diagrams, one with the
photon attached to the fermion and the other to the scalar. Also in
this case it has been calculated long ago~\cite{Gunion:1989in} and it can
also be derived from the results of Ref.~\cite{Chiu:2014oma}. The result is again
negative, so that the addition of $S_2$ is not useful to explain the $a_\mu$ discrepancy.\\

\noindent $\mathbf{H_2 \sim (1,2,-1/2)}$

In the case of adding a replica of the Higgs doublet, the issue is a
bit more subtle, since the neutral component, which
couples to the muons, can adquire a vacuum expectation value ($vev$). 

In Ref.~\cite{Freitas:2014pua} they assume that the new scalar does
not develope any $vev$ contributing to the fermion masses. In this case they
show that in principle there could be a positive contribution to
$a_\mu$ able to explain the current discrepancy; however, this is now excluded by bounds coming from four-fermion
interactions and searches for neutralino resonances already at LEP.

On the other hand the situation is even more involved in the case the
scalar takes $vev$, since a choice can be done regarding which of the
two scalars contribute to various fermion masses. In
Ref.~\cite{Broggio:2014mna} they consider four two-Higgs-doublet
models and conclude that in one of them $\Delta_\mu$ can be
obtained, thanks to the enhanced two-loops contribution of the
pseudoscalar Higgs. In order for this to be realized, all the scalars
must be lighter than $\sim 200$~GeV, and therefore this option is testable
at the LHC. For details, we remand to the mentioned paper.\\

\noindent $\mathbf{\Delta \sim (1,3,0)}$

In this case only the two charged components of the triplet contribute
to $a_\mu$\footnote{Indeed the neutral component only couples to
neutrinos~\cite{Gunion:1989in,Abada:2007ux}.} and the final result is essentially
the sum of the results previously mentioned for the singly- and
 doubly-charged scalar. Therefore it is negative and this
minimal extension of the SM can not explain the experimental result.\\

\noindent {\bf Leptoquarks}

The possibility of explaining the $a_\mu$ discrepancy by adding a
single leptoquark (LQ) has been discussed with great details in
Ref.~\cite{Chakraverty:2001yg}, where they have concluded that this is
indeed possible. In fact in this case, if the LQ has both left-handed
and right-handed couplings to the muon, the
contribution to $a_\mu$ is enhanced by the mass of the quark circulating in
the loop, which must be of up-type. From our list of five leptoquarks,
it can be shown that only $S_c^{1/3}$ and $D_c^{7/6}$ respect this
condition and, therefore, they are good candidates to solve the $(g-2)_\mu$
puzzle.~\footnote{These scalar LQs have been recently considered in
Ref.~\cite{Queiroz:2014pra}. Among other things, they have
shown that $D_c^{7/6}$ does not give rise to dangerous proton decay.}

In order to have an enhancement, either the top or the charm have to
circulate in the loop. The contribution to $a_\mu$ will be
proportional to $\frac{\lambda^{LQ \mu t(c)}_L \lambda^{LQ \mu
    t(c)}_R}{M_{LQ}^2}$ when the top (charm) is considered. Then a relevant
question arises, i.e. which are the bounds on these quantities.

Let's start with the top. So far, there are no bounds coming from
colliders on a leptoquark decaying into top and muon. However, in the case
of $S_c^{1/3}$, bounds come from the decay into $b\nu$, giving
$M_{S_c^{1/3}}>620$~GeV~\cite{Aad:2013ija}. Even if these
recent limits are stronger then the ones considered in
Ref.~\cite{Chakraverty:2001yg}, they are not strong enough to rule
out this possibility as an explanation of the $a_\mu$
anomaly. Indeed, the couplings remains perturbative ($\lambda < 1\
(4\pi)$) for masses smaller than 40 (500)~TeV. 

An example of LQ decaying
in this way is the $\tilde{b}_R$ of the
supersymmetric model of Ref.~\cite{Riva:2012hz}. Since in this model
the coupling is fixed to be the Yukawa coupling, if we want to explain the $a_\mu$ discrepancy we obtain a
prediction for the sbottom mass, which has to be around
500~GeV. This seems to be excluded from the quoted mass limit which,
however, is obtained assuming a unit branching ratio. Actually, if the
branching fraction is reduced to $60\%$, the bound is reduced to
520~GeV~\cite{Aad:2013ija}. Therefore the $\tilde{b}_R$ of this model
could still be the responsible for the disagreement between the
measured and predicted values of the muon anomalous magnetic
moment. This enforces the need of searches for 
final states with tops and muons. 

On the other hand, if the scalar LQ couples to muon and charm, the
stringent bounds coming from LHC searches with muons and jets in the final
state apply and we have $M_{LQ}>1070$~GeV (if
Br=1)~\cite{LQ2ndgen}. However, even if with this strong bound on the LQ
mass larger couplings are needed, they are still perturbative, so that
also a charm circulating in the loop can give a sizeable contribution
to $a_\mu$. Indeed, in this case, the requirement of perturbativity
for the LQ coupling gives the following constraint on the LQ mass:
$M_{LQ}<4\ (60)$~TeV (for $\lambda<1\ (4\pi)$), making the search at
the LHC even more intriguing.

\section{Conclusions}

In this work we have completed the analysis of the contributions to
the muon anomalous magnetic moment for all the one-particle extensions
of the SM (scalars and fermions). In particular we have performed the
calculation in the case of a colourless fermion, SU(2) doublet with
hypercharge $-3/2$, which was absent in the literature, obtaining a
negative contribution to $a_\mu$. From the analysis performed we can therefore conclude
that the addition of a single fermion to the SM cannot explain the
measuread discrepancy. On the other hand, {\it the addition of a single
scalar could be the responsible of the discrepancy, if
the new scalar is a second Higgs doublet or one of the two scalar
leptoquarks $S_c^{1/3}$ or $D_c^{7/6}$}. 

If the solution of the muon anomalous magnetic moment puzzle comes
from a single BSM particle, this could be tested in the next run of
the LHC. In the case of a second
Higgs doublet, the corresponding particles should be lighter than
200~GeV and therefore one expects that the available
parameter space will be covered by the next LHC searches. On the other hand, if
$\Delta_\mu$ is generated by a leptoquark, mass bounds coming from the
perturbativity requirement are not very
stringent and in principle it could lie beyond the LHC
reach. However, we have shown a particular model where this can happen
and the LQ mass should be just behind the corner. Since leptoquarks
have peculiar signatures at the LHC, we suggest to enforce these
searches and, in particular, to look for leptoquarks decaying into
top and muons, a channel not yet considered at the LHC.

\section*{Acknowledgements}

We would like to thank Michele Frigerio for early discussions about
this project and Lorenzo Calibbi for comments on the first version of
this manuscript. This research was partially supported
by the Marie Curie CIG program, project number PCIG13-GA-2013-618439, and
by the Swiss National Science Foundation (SNF) under contract 200020-146644.

\section*{Appendix}

In the interaction basis where the Lagrangian of Eq.~(\ref{Lagrangian})
is defined, the mass term for the singly charged leptons is given by
\begin{equation}
\label{Massterm}
-\begin{pmatrix}
\bar{\ell}_L & \bar{\chi}_L
\end{pmatrix} 
\begin{pmatrix}
\frac{yv}{\sqrt{2}} & 0 \\[1.5mm]
\frac{\lambda_D v}{\sqrt{2}} & M_D
\end{pmatrix} 
\begin{pmatrix}
\ell_R \\[1mm]
\chi_R
\end{pmatrix} 
+h.c. \ .
\end{equation}
The mass matrix is diagonalized with a bi-unitary transformation:
\begin{equation}
\begin{pmatrix}
\ell_{L,R} \\[2mm]
\chi_{L,R}
\end{pmatrix}
=V_{L,R} \ 
\begin{pmatrix}
\ell^m_{L,R} \\[2mm]
\chi^m_{L,R}
\end{pmatrix}
\end{equation}
where the superscript $m$ indicates the mass basis, and the unitary
matrices performing the rotation are given by:
\begin{equation}
V_{L}=
\begin{pmatrix}
\mathbb{1} & \frac{y\lambda_{D}^{\dagger} v^2}{2M_D^{2}} \\[2mm]
-\frac{\lambda_{D} yv^{2}}{2M_D^{2}} & 1
\end{pmatrix}
\qquad \text{and} \qquad
V_{R}=
\begin{pmatrix}
\mathbb{1}-\frac{\lambda_D^\dagger\lambda_{D} v^2}{4M_D^{2}} &  \frac{\lambda_D^{\dagger} v}{\sqrt{2}M_D} \\[2mm]
-\frac{\lambda_{D} v}{\sqrt{2}M_D} &
1-\frac{\lambda_{D}\lambda_D^{\dagger} v^2}{4M_D^{2}}
\end{pmatrix}\ .
\end{equation}
In the mass basis, the Lagrangian turns out to be:
\begin{eqnarray}
\mathcal{L}_{\text{em}}&=& -e\left(\bar{l}^{m}\gamma^{\mu}l^m+\bar{\chi}^{m}\gamma^{\mu}\chi^m+2\bar{\Psi}\gamma^{\mu}\Psi\right)A_{\mu} \\[2mm]
\mathcal{L}_{\text{kin}}&=&i\left(\bar{l}^{m}\slashed{\partial} l^{m}+\bar{\chi}^{m}\slashed{\partial}\chi^{m} + \bar{\Psi}\slashed{\partial}\Psi\right) \ 
\\[2mm]
\mathcal{L}_{\text{NC}}^{\nu \Psi} &=&
\frac{g}{2\cos\theta_w}\left (
\bar{\nu}\gamma^{\mu}P_L\nu+\bar{\Psi}\gamma^{\mu}\left(-1+4\sin^{2}\theta_w\right)\Psi
\right )Z_{\mu}\\
\mathcal{L}_{\text{NC}}^{l\chi} &=&
\begin{pmatrix}
\bar{l}^{m} & \bar{\chi}^{m}
\end{pmatrix}
\gamma^{\mu}
\left(g^{Z}_{R}P_{R}+g^{Z}_{L}P_{L}\right)
\begin{pmatrix}
l^{m} \\[2mm]
\chi^{m}
\end{pmatrix}
Z_{\mu}
\\
\mathcal{L}_{\text{CC}}^{\Psi}&=&\ \bar{\Psi}\gamma^{\mu}W_{\mu}^{-}g^{W}_{\Psi l}l^{m}+\bar{\Psi}\gamma^{\mu}W_{\mu}^{-}g^{W}_{\Psi\chi}\chi^{m}+h.c. \\[2mm]
\mathcal{L}_{\text{CC}}^{\nu}&=&\bar{\nu}\gamma^{\mu}W_{\mu}^{+}g^{W}_{\nu l}l^{m}+\bar{\nu}\gamma^{\mu}W_{\mu}^{+}g^{W}_{ \nu\chi}\chi^{m} +h.c.
\\
\mathcal{L}_{h}&=&-
\begin{pmatrix}
\bar{l}^{m} & \bar{\chi}^{m}
\end{pmatrix}
\left(g^{h}_{R}P_{R}+g^{h}_{L}P_{L}\right)
\begin{pmatrix}
l^{m} \\[2mm]
\chi^{m}
\end{pmatrix}
h
\end{eqnarray}
\begin{eqnarray}
\mathcal{L}_{\phi}&=&\bar{\Psi}g^{\phi}_{\Psi l}l^{m}\phi^{-}+\bar{\Psi}g^{\phi}_{\Psi\chi}\chi^{m}\phi^{-}-\bar{l}^{m}g^{\phi}_{l\nu}\phi^{-}\nu-\bar{\chi}^{m}g^{\phi}_{\chi\nu}\nu\phi^{-}+h.c.
\\
\mathcal{L}_{\eta}&=&-
\begin{pmatrix}
\bar{l}^{m} & \bar{\chi}^{m}
\end{pmatrix}
\left(g^{\eta}_{R}P_{R}+g^{\eta}_{L}P_{L}\right)
\begin{pmatrix}
l^{m} \\[2mm]
\chi^{m}
\end{pmatrix}
\eta
\end{eqnarray}
where:
\begin{subequations}
\begin{align}
&g^{Z}_{R} = 
\begin{pmatrix}
g^{Z}_{R_{ll}} & g^{Z}_{R_{l\chi}} \\[2mm]
g^{Z}_{R_{\chi l}} & g^{Z}_{R_{\chi\chi}}
\end{pmatrix}
=\frac{g}{\cos\theta_{w}}
\begin{pmatrix}
\left(1-\cos^{2}\theta_{w}\right)\mathbb{1}+\frac{v^{2}\lambda_D^\dagger\lambda_D}{4M_D^{2}} & -\frac{v\lambda_D^\dagger}{2\sqrt{2}M_D} \\[2mm]
-\frac{v\lambda_D}{2\sqrt{2}M_D} & \frac{3}{2}-\cos^{2}\theta_{w}-\frac{v^{2}\lambda_D\lambda_D^\dagger}{4M^{2}_{D}}
\end{pmatrix}
\\[1mm]
&g^{Z}_{L}=
\begin{pmatrix}
g^{Z}_{L_{ll}} & g^{Z}_{L_{l\chi}} \\[2mm]
g^{Z}_{L_{\chi l}} & g^{Z}_{L_{\chi\chi}}
\end{pmatrix}
=\frac{g}{\cos\theta_{w}}
\begin{pmatrix}
\mathbb{1}\left(\frac{1}{2}-\cos^{2}\theta_{w}\right) & -\frac{v^{2}y\lambda_D^\dagger}{2M_{D}^{2}} \\[2mm]
-\frac{v^{2}\lambda_D y}{2M_{D}^{2}} & \frac{3}{2}-\cos^{2}\theta_{w}
\end{pmatrix}
\\[1mm]
&g^{h}_{R}=
\begin{pmatrix}
g^{h}_{R_{ll}} & g^{h}_{R_{l\chi}} \\[2mm]
g^{h}_{R_{\chi l}} & g^{h}_{R_{\chi\chi}}
\end{pmatrix}
=\frac{1}{\sqrt{2}}
\begin{pmatrix}
y-\frac{3y\lambda_D^\dagger\lambda_D v^{2}}{4M_{D}^{2}} &
\frac{y\lambda_D^\dagger v}{\sqrt{2}M_{D}} \\[2mm]
 \lambda_D+\frac{v^{2}\lambda_D y^{2}}{2M_D^{2}}-\frac{v^{2}\lambda_D\lambda_D^\dagger\lambda_D}{4M_D^{2}}           &
 \frac{\lambda_D\lambda_D^\dagger v}{2M_D}
\end{pmatrix}
\\[1mm]
&g^{h}_{L}=
\begin{pmatrix}
g^{h}_{L_{ll}} & g^{h}_{L_{l\chi}} \\[2mm]
g^{h}_{L_{\chi l}} & g^{h}_{L_{\chi\chi}}
\end{pmatrix}
=\frac{1}{\sqrt{2}}
\begin{pmatrix}
y-\frac{3\lambda_D^\dagger\lambda_D yv^{2}}{4M_D^{2}} & 
\lambda_D^\dagger+\frac{v^{2}y^{2}\lambda_D^\dagger}{2M_D^{2}}-\frac{v^{2}\lambda_D^\dagger\lambda_D\lambda_D^\dagger}{4M_D^{2}} \\[2mm]
\frac{\lambda_D yv}{\sqrt{2}M_D} & 
\frac{\lambda_D\lambda_D^\dagger v}{2M_D}
\end{pmatrix}
\\[1mm]
&g^{\eta}_{R}=
\begin{pmatrix}
g^{\eta}_{R_{ll}} & g^{\eta}_{R_{l\chi}} \\
g^{\eta}_{R_{\chi l}} & g^{\eta}_{R_{\chi\chi}}
\end{pmatrix}
=\frac{i}{\sqrt{2}}
\begin{pmatrix}
\frac{y\lambda_D^\dagger\lambda^{ D} v^{2}}{4M_D^{2}}+y & \frac{y\lambda_D^\dagger v}{\sqrt{2}M_D} \\[2mm]
-\lambda_D+\frac{\lambda_D
  y^{2}v^{2}}{2M_D^{2}}+\frac{v^{2}\lambda_D\lambda_D^\dagger\lambda_D}{4M_D^{2}}
& -\frac{\lambda_D\lambda_D^\dagger v}{\sqrt{2}M_D}
\end{pmatrix}
\\[1mm]
&g^{\eta}_{L}=
\begin{pmatrix}
g^{\eta}_{L_{ll}} & g^{\eta}_{L_{l\chi}} \\
g^{\eta}_{L_{\chi l}} & g^{\eta}_{L_{\chi\chi}}
\end{pmatrix}
=\frac{i}{\sqrt{2}}
\begin{pmatrix}
-\frac{\lambda_D^\dagger\lambda_D yv^{2}}{4M_D^{2}}-y & +\lambda_D^\dagger-\frac{y^{2}v^{2}\lambda_D^\dagger}{2M_D^{2}}-\frac{v^{2}\lambda_D^\dagger\lambda_D\lambda^{ D^{\dagger}}}{4M_D^{2}} \\[2mm]
-\frac{\lambda_D yv}{\sqrt{2}M_D} & \frac{\lambda_D\lambda_D^{\dagger}v}{\sqrt{2}M_D}
\end{pmatrix} \\[2mm]
&\begin{cases}
g^{W}_{\nu\chi} = \frac{g}{\sqrt{2}}\frac{v^{2}y\lambda_D^\dagger}{2M_D^{2}}P_{L} \\[2mm]
g^{W}_{\nu l}=\frac{g}{\sqrt{2}}\mathbb{1}P_{L}
\end{cases}
\qquad \qquad \qquad \ 
\begin{cases}
g^{W}_{\Psi l} = \frac{g}{\sqrt{2}}\left[-\frac{v^{2}\lambda_D y}{2M_D^{2}}P_{L}-\frac{v\lambda_D}{\sqrt{2}M_D}P_{R}\right] \\[2mm]
g^{W}_{\Psi\chi} = \frac{g}{\sqrt{2}}\left[P_{L}+\left(1-\frac{v^{2}\lambda_D\lambda_D^\dagger}{4M_D^{2}}\right)P_{R}\right] \
\end{cases}
\\[2mm]
&
\begin{cases}
g^{\phi}_{\Psi l}= \ \lambda_D\left(\mathbb{1}-\frac{v^{2}\lambda_D^\dagger\lambda_D}{4M_D^{2}}\right)P_{R} \\[2mm]
g^{\phi}_{\Psi\chi}= \ \frac{\lambda_D\lambda_D^\dagger v}{\sqrt{2}M_D}P_{R}
\end{cases}
\qquad
\begin{cases}
g^{\phi}_{l\nu}=\left(y-\frac{v^{2}\lambda_D^\dagger\lambda_D y}{4M_D^{2}}\right)P_{L} \\[2mm]
g^{\phi}_{\chi\nu}= \ \frac{\lambda_D vy}{\sqrt{2}M_D}P_{L} \ .
\end{cases}
\end{align}
\end{subequations}
Computing the diagrams in Fig.~\ref{fig:diagrams} we find the following
results:
\begin{eqnarray}
a_{\mu}^{Z\chi}&=&\frac{G_{F}m_{\mu}^{2}}{24\sqrt{2}\pi^{2}}\frac{v^{2}\left|\lambda_{D\mu}\right|^{2}}{M_{D}^{2}}\frac{8-3k_{Z}-12k_{Z}^2+7k_{Z}^3+6\left(6k_{Z}-1-4k_{Z}^2\right)\log
  k_{Z}}{2\left(k_{Z}-1\right)^4} \\
a_{\mu}^{Z\mu}&=&\frac{G_{F}m_{\mu}^{2}}{24\sqrt{2}\pi^{2}}\left[\left(3-4\cos^{2}\theta_{w}\right)^{2}-5-\frac{2\left|\lambda_{D\mu}\right|^{2}v^{2}}{M_{D}^{2}}\left(1+2\cos^{2}\theta_{w}\right)\right]\\
a_{\mu}^{\eta\mu}&=&\frac{11
  G_{F}m_{\mu}^2}{24\sqrt{2}\pi^{2}}\frac{m_{\mu}^2}{M_{Z}^{2}}\left(1+\frac{v^{2}\left|\lambda_{D\mu}\right|^{2}}{2M_{D}^{2}}\right)
\end{eqnarray}
\begin{eqnarray}
a_{\mu}^{\eta\chi}&=&
\frac{ G_F m_\mu^{2}}{16\sqrt{2}\pi^{2}} \
\frac{\left|\lambda_{D\mu}\right|^{2}v^{2}}{M_D^{2}} \ 
\frac{-5k_Z^{4}+24k_Z^{3}-39k_Z^{2}+20k_Z+6k_Z\log(k_Z)}{6(k_Z-1)^{4}}\\
a_{\mu}^{h\chi}&=&\frac{G_{F} m_{\mu}^{2}}{16\sqrt{2}\pi^{2}} \
\frac{\left|\lambda_{D\mu}\right|^{2}v^{2}}{M_D^{2}} \
\frac{7k_{h}^{4}-36k_{h}^{3}+45k_{h}^{2}-16k_{h}+6k_{h}(3k_{h}-2)\log(k_{h})}{6(k_{h}-1)^{4}}
\\
a_{\mu}^{h\mu} &=& -\frac{7
  G_{F}m_{\mu}^2}{24\sqrt{2}\pi^{2}} \frac{m_{\mu}^2}{m_{h}^{2}}
\left(1-\frac{3    \left|\lambda_{D\mu}\right|^{2}}{2M_{D}^{2}}\right)\\
a_{\mu_{1}}^{W\nu}& =& \frac{7 G_F m_\mu^2}{24\sqrt{2}\pi^2} \\
a_{\mu_{1}}^{W\Psi}& =&-\frac{G_{F} m_{\mu}^{2}}{8\sqrt{2}\pi^{2}}
\frac{v^{2}\left|\lambda_{D\mu}\right|^{2}}{M_{D}^{2}}
\frac{-11+57k_{W}-69k_{W}^{2}+23k_{W}^{3}-6k_{W}^{2}\left(-5+3k_{W}\right)\log
  k_{W}}{6\left(k_{W}-1\right)^{4}}\\
a_{\mu_{2}}^{W\Psi} &=
&\frac{G_{F} m_{\mu}^{2}}{2\sqrt{2}\pi^{2}}\frac{v^{2}\left|\lambda_{D\mu}\right|^{2}}{M_{D}^{2}}\frac{8-3k_{W}-12k_{W}^{2}+7k_{W}^{3}-6\left(1-6k_{W}+4k_{W}^{2}\right)\log
  k_{W}}{6\left(k_{W}-1\right)^{4}}\\
a_{\mu}^{\phi\nu_{\mu}}&=&-\frac{G_{F}m_{\mu}^2}{24\sqrt{2}}\frac{m_{\mu}^2}{M_{W}^{2}}
\left(1-\frac{\left|\lambda_{D\mu}\right|^{2}v^{2}}{2M_{D}^{2}}\right)\\
a_{\mu_{1}}^{\phi\Psi}&=&\frac{G_{F} m_{\mu}^{2}}{8\sqrt{2}\pi^{2}}\frac{v^{2}\left|\lambda_{D\mu}\right|^{2}}{M_{D}^{2}}k_{W}\frac{\left(k_{W}-1\right)\left[-1+k_{W}\left(5+2k_{W}\right)\right]-6k_{W}^{2}\log
  k_{W}}{6\left(k_{W}-1\right)^{4}}\\
a_{\mu_{2}}^{\phi\Psi}&=&\frac{G_{F} m_{\mu}^{2}}{4\sqrt{2}\pi^{2}} \
\frac{v^{2}\left|\lambda_{D\mu}\right|^{2}}{M_{D}^{2}} \
k_{W}\frac{2+3k_{W}-6k_{W}^{2}+k_{W}^{3}+6k_{W}\log
  k_{W}}{6\left(k_{W}-1\right)^{4}} \\
a_{\mu}^{W\phi}&=&\frac{G_{F} m_{\mu}^{2}}{16\sqrt{2}\pi^{2}}\left(1-\frac{v^{2}\left|\lambda_{D\mu}\right|^{2}}{4M_{D}^{2}}\right)
\\
a_{\mu}^{W\phi\Psi}&=&\frac{G_{F} m_{\mu}^{2}}{16\sqrt{2}\pi^{2}} \ \frac{v^{2}\left|\lambda_{D\mu}\right|^{2}}{M_{D}^{2}} \ \frac{-1+(4-3k_{W})k_{W}+2k_{W}^{2}\log k_{W}}{2\left(k_{W}-1\right)^{3}}
\end{eqnarray}
where $k_{Z}=M_{D}^{2}/M_{Z}^{2}$, $k_{W}=M_{D}^{2}/M_{W}^{2}$ and
$k_{h}=M_{D}^{2}/M_{h}^{2}$. Summing all these contributions together
(and twice the last two since each of them corresponds to two
diagrams) we obtain the result shown in Eq.\eqref{amuD1}.



\begin{thebibliography}{99}
%

\bibitem{Bennett:2006fi}
  G.~W.~Bennett {\it et al.}  [Muon G-2 Collaboration],
  Phys.\ Rev.\ D {\bf 73} (2006) 072003
  [hep-ex/0602035].

\bibitem{Jegerlehner:2009ry}
  F.~Jegerlehner and A.~Nyffeler,
  Phys.\ Rept.\  {\bf 477} (2009) 1
  [arXiv:0902.3360 [hep-ph]].

\bibitem{Davier:2003pw}
  M.~Davier, S.~Eidelman, A.~Hocker and Z.~Zhang,
  Eur.\ Phys.\ J.\ C {\bf 31} (2003) 503
  [hep-ph/0308213].


\bibitem{passeraetal}
  M.~Passera, W.~J.~Marciano and A.~Sirlin,
  Phys.\ Rev.\ D {\bf 78} (2008) 013009
  [arXiv:0804.1142 [hep-ph]];
  M.~Passera, W.~J.~Marciano and A.~Sirlin,
  AIP Conf.\ Proc.\  {\bf 1078} (2009) 378
  [arXiv:0809.4062 [hep-ph]];
  M.~Passera, W.~J.~Marciano and A.~Sirlin,
  Chin.\ Phys.\ C {\bf 34} (2010) 735
  [arXiv:1001.4528 [hep-ph]].

\bibitem{fermilab-exp}
  R.~M.~Carey, K.~R.~Lynch, J.~P.~Miller, B.~L.~Roberts, W.~M.~Morse, Y.~K.~Semertzides, V.~P.~Druzhinin and B.~I.~Khazin {\it et al.},
  FERMILAB-PROPOSAL-0989.

\bibitem{jparc}
http://j-parc.jp/researcher/Hadron/en/pac\_0907/pdf/LOI\_Saito.pdf


\bibitem{Queiroz:2014zfa}
  F.~S.~Queiroz and W.~Shepherd,
  Phys.\ Rev.\ D {\bf 89} (2014) 095024
  [arXiv:1403.2309 [hep-ph]].


\bibitem{typeI}
P. Minkowski, Phys.\ Lett.\ B 67 421 (1977);
M. Gell-Mann, P. Ramond and R. Slansky,
in {\it Supergravity}, edited by P. van Nieuwenhuizen and D. Freedman,
(North-Holland, 1979), p.~315;
T. Yanagida, in {\it Proceedings of the Workshop on the Unified Theory
and the Baryon Number in the Universe}, edited by O.~Sawada and
A. Sugamoto (KEK Report No.~79-18, Tsukuba, 1979), p.~95;
R.N.~Mohapatra and G.~Senjanovi\'{c}, Phys. Rev. Lett. {\bf 44} (1980) 912.

\bibitem{typeIII}
  R.~Foot, H.~Lew, X.~G.~He and G.~C.~Joshi,
  Z.\ Phys.\  C {\bf 44}, 441 (1989).

\bibitem{Delgado:2011iz}
  A.~Delgado, C.~Garcia Cely, T.~Han and Z.~Wang,
  Phys.\ Rev.\ D {\bf 84} (2011) 073007
  [arXiv:1105.5417 [hep-ph]].

\bibitem{Ma:2014zda}
  T.~Ma, B.~Zhang and G.~Cacciapaglia,
  Phys.\ Rev.\ D {\bf 86} (2012) 035001
  [arXiv:1404.2375 [hep-ph]].

\bibitem{delAguila:2008pw}
  F.~del Aguila, J.~de Blas and M.~Perez-Victoria,
  Phys.\ Rev.\ D {\bf 78} (2008) 013010
  [arXiv:0803.4008 [hep-ph]];
  J.~de Blas,
  EPJ Web Conf.\  {\bf 60} (2013) 19008
  [arXiv:1307.6173 [hep-ph]].


\bibitem{Chakraverty:2001yg}
  D.~Chakraverty, D.~Choudhury and A.~Datta,
  Phys.\ Lett.\ B {\bf 506} (2001) 103
  [hep-ph/0102180].

\bibitem{g-2-EW}
  R.~Jackiw and S.~Weinberg,
  Phys.\ Rev.\  D {\bf 5} (1972) 2396;
  I.~Bars and M.~Yoshimura,
  Phys.\ Rev.\  D {\bf 6} (1972) 374;
  G.~Altarelli, N.~Cabibbo and L.~Maiani,
  Phys.\ Lett.\  B {\bf 40} (1972) 415;
  W.~A.~Bardeen, R.~Gastmans and B.~Lautrup,
  Nucl.\ Phys.\  B {\bf 46} (1972) 319;
  K.~Fujikawa, B.~W.~Lee and A.~I.~Sanda,
  Phys.\ Rev.\  D {\bf 6} (1972) 2923.



\bibitem{pdg}
K.A. Olive et al. (Particle Data Group), Chin.\ Phys.\ C,\ 38, 090001 (2014).

\bibitem{Biggio:2008in}
  C.~Biggio,
  Phys.\ Lett.\ B {\bf 668} (2008) 378
  [arXiv:0806.2558 [hep-ph]].

\bibitem{Freitas:2014pua}
  A.~Freitas, J.~Lykken, S.~Kell and S.~Westhoff,
  JHEP {\bf 1405} (2014) 145
  [arXiv:1402.7065 [hep-ph]].

\bibitem{Chiu:2014oma}
  W.~C.~Chiu, C.~Q.~Geng and D.~Huang,
  arXiv:1409.4198 [hep-ph].

\bibitem{Kannike:2011ng}
  K.~Kannike, M.~Raidal, D.~M.~Straub and A.~Strumia,
  JHEP {\bf 1202} (2012) 106
   [Erratum-ibid.\  {\bf 1210} (2012) 136]
  [arXiv:1111.2551 [hep-ph]].

\bibitem{Dermisek:2013gta}
  R.~Dermisek and A.~Raval,
  Phys.\ Rev.\ D {\bf 88} (2013) 013017
  [arXiv:1305.3522 [hep-ph]].


\bibitem{CoarasaPerez:1995wa}
  J.~A.~Coarasa Perez, A.~Mendez and J.~Sola,
  Phys.\ Lett.\ B {\bf 374} (1996) 131
  [hep-ph/9511297].

\bibitem{Gunion:1989in}
  J.~F.~Gunion, J.~Grifols, A.~Mendez, B.~Kayser and F.~I.~Olness,
  Phys.\ Rev.\ D {\bf 40} (1989) 1546.

\bibitem{Broggio:2014mna}
  A.~Broggio, E.~J.~Chun, M.~Passera, K.~M.~Patel and S.~K.~Vempati,
  arXiv:1409.3199 [hep-ph].

\bibitem{Abada:2007ux}
  A.~Abada, C.~Biggio, F.~Bonnet, M.~B.~Gavela and T.~Hambye,
  JHEP {\bf 0712} (2007) 061
  [arXiv:0707.4058 [hep-ph]].

\bibitem{Queiroz:2014pra}
  F.~S.~Queiroz, K.~Sinha and A.~Strumia,
  arXiv:1409.6301 [hep-ph].

\bibitem{Aad:2013ija}
  G.~Aad {\it et al.}  [ATLAS Collaboration],
  JHEP {\bf 1310} (2013) 189
  [arXiv:1308.2631 [hep-ex]].

\bibitem{Riva:2012hz}
  F.~Riva, C.~Biggio and A.~Pomarol,
  JHEP {\bf 1302} (2013) 081
  [arXiv:1211.4526 [hep-ph]].


\bibitem{LQ2ndgen}
CMS-PAS-EXO-12-042.

\end{thebibliography}
\end{document}